\documentclass{ws-procs975x65}

\begin{document}

\title{Isobar Model for Photoproduction of $K^+\Sigma^0$ and $K^0\Sigma^+$ on
the Proton.}

\author{T. Mart}

\address{Departemen Fisika, FMIPA, Universitas Indonesia, Depok 16424, 
  Indonesia}


\maketitle

\abstracts{Kaon photoproductions on the proton $\gamma p\to K^+\Sigma^0$ and 
$\gamma p\to K^0\Sigma^+$ have been simultaneously analyzed by using isobar
models and new SAPHIR data. The result shows that isobar models such as KAON MAID
require more resonances in order to explain the data.}

\section{Introduction}
It has been widely known that pion nucleon scattering provides an excellent tool
for investigation of nucleon resonances. Indeed, this process has become 
a longstanding source of information on nucleon resonances properties, 
such as their masses, decay widths, electromagnetic and hadronic coupling 
constants, as well as their spins. However, the exclusive use of pion nucleon
scattering would bias the information on the existence of certain resonances.
Modern quark model studies predict a much richer resonance spectrum\,%
\cite{NRQM,capstick94} than has been observed in $\pi N\to \pi N$ 
scattering experiments. Where have all these resonances gone? The answer
also comes from these quark models. They have suggested that those "missing" 
resonances may couple strongly to other channels, such as the $K \Lambda$ 
and $K \Sigma$ channels\,\cite{capstick98} or other final states involving 
vector mesons. Since performing kaon-nucleon or hyperon-nucleon scattering 
experiments is a daunting task, kaon photoproduction on the nucleon 
appears as the best solution.

The new generation of electron and photon facilities have paved the way to
kaon photoproduction experiments of unprecedented accuracies. Such experiments
have been performed at {\small ELSA} in Bonn, {\small JLab} in Newport News, 
{\small SPring8} in Osaka, and {\small ESRF} in Grenoble. In this paper
we will only focus on the newly published data by {\small SAPHIR} collaboration 
of {\small ELSA}\,\cite{Lawall:2005np,Glan03b} for the $\gamma p\to K^+\Sigma^0$ 
and $\gamma p\to K^0\Sigma^+$ channels. These data could be very interesting,
since they indicate that isobar models (e.g., {\small KAON MAID}\,\cite{kaonmaid})
require more resonances in order to reproduce them.

\section{Isobar Model}
We start with the {\small KAON MAID} model\,\cite{kaonmaid}. The model consists
of the standard Born terms, along with the $K^*$ and $K_1$ exchanges as 
the background part, 
while the resonance part includes the isospin 1/2 nucleon resonances $S_{11}(1650)$, 
$P_{11}(1710)$, and $P_{13}(1720)$, as well as the isospin 3/2 deltas $S_{31}(1900)$ 
and $P_{31}(1910)$. A brief discussion of the model can be found in 
Refs.\,\cite{Mart:1999ed,Mart:2000}.

\section{Result and Discussion}
\begin{figure}[!t] 
\centerline{\epsfxsize=5in\epsfbox{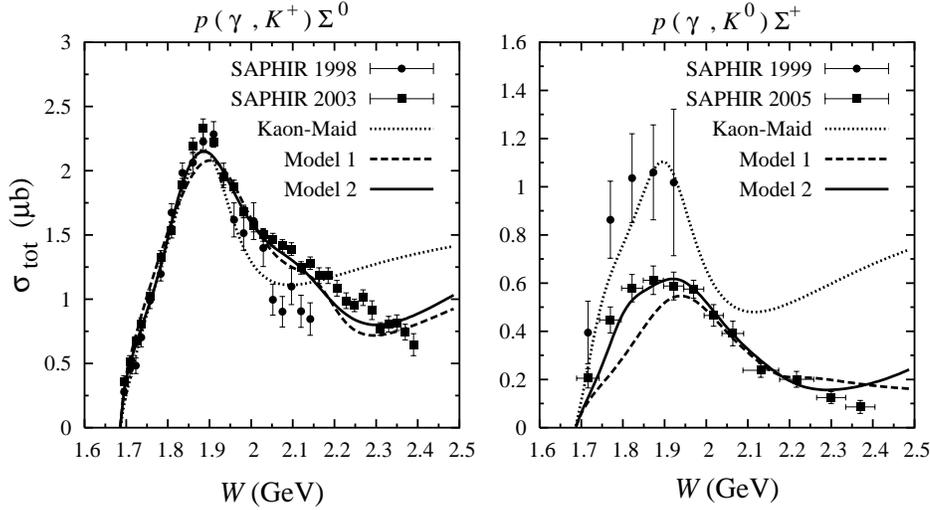}}   
\caption{Total cross section for $\gamma p\to K^+\Sigma^0$ and the
$\gamma p\to K^0\Sigma^+$ channels. Experimental data are taken from
Refs.\,\protect\cite{Lawall:2005np,Goers:1999sw,Tran:1998qw,Glan03b}.
\label{total}}
\end{figure}

As shown by the two panels of Fig.\,\ref{total} {\small KAON MAID} obviously
cannot reproduce the new data, especially in the $K^0\Sigma^+$ channel. This
shortcoming is understandable, since {\small KAON MAID} was fitted to previous 
{\small SAPHIR} data. From this figure it is also clear that
the new data do not exhibit a certain resonance structure and, therefore, 
an arbitrarily inclusion of new resonances in the model is not
advocated. However, a close inspection to the $K^+\Sigma^0$ total cross section 
reveals that a significant discrepancy between the prediction of {\small KAON MAID} and 
new data exists at $W\simeq 2.1$ GeV. This discrepancy originates from the previous
data, where no indication of resonances required by the model at this energy as well
as due to the large error-bars. 

\begin{table}[bt]
\tbl{Masses and widths of resonances from the three models.}
{\footnotesize
\begin{tabular}{@{}lcrrr@{}}
\hline\\[-1.8ex]
Resonance & Mass or & Original & Model 1 & Model 2 \\
 & Width & value \cite{pdg} & & \\
\hline\\[-1.4ex]
$S_{11}(1650)$ & $M$ & 1650 & 2167 & 1795 \\
          & $\Gamma$ &  150 &  186 &  158 \\
               & $M$ &  $-$ &  $-$ & 2112 \\
          & $\Gamma$ &  $-$ &  $-$ &  400 \\
$P_{11}(1710)$ & $M$ & 1710 & 1690 & 1680 \\
          & $\Gamma$ &  100 &  100 &  100 \\
$P_{13}(1720)$ & $M$ & 1720 & 2133 & 2141 \\
          & $\Gamma$ &  150 &  256 &  279 \\
$S_{31}(1900)$ & $M$ & 1900 & 1920 & 1900 \\
          & $\Gamma$ &  200 &  355 &  329 \\
$P_{31}(1910)$ & $M$ & 1910 & 1936 & 1800 \\
          & $\Gamma$ &  250 &  399 &  400 \\
\hline\\[-2.2ex]
$\chi^2/N$ && 4.14 & 2.44 & 1.76 \\
\hline
\end{tabular}\label{extract} }
\end{table}

To investigate this phenomenon we refit the original {\small KAON MAID} 
coupling constants by including  only new data in our database. 
The obtained $\chi^2$ per degrees of freedom is 
shown in the third column of Table\,\ref{extract}, which 
is clearly far from satisfactory. In the second step,
we allow the masses and widths of nucleon resonances to be determined by the fit.
As shown by the fourth column of Table\,\ref{extract}, the result is quite
interesting. Besides significantly reducing $\chi^2$, the fit shifts 
both $S_{11}$ and $P_{13}$ masses to higher values (2167 MeV and 2133 MeV,
respectively), while other resonances seem to be more stable. From the
experience in the multipoles study of kaon photoproduction\,\cite{Mart:2004ug}, 
we suspect that such behavior could be an indication for the existence of similar  
resonances, but with relatively different masses. Therefore, in the next
step we put two $S_{11}$ resonances and leave their masses and widths to 
be determined by the fit. As shown by the fifth column of the same Table, we obtain 
two $S_{11}$s with masses 1795 MeV and 2112 MeV, which is seemingly 
to support the finding of Ref.\,\cite{Mart:2004ug}.
The result clearly indicates that more resonances are 
required to explain kaon photoproduction process, a point which should 
be addressed in future studies.

\begin{figure}[!t] 
\centerline{\epsfxsize=5.1in\epsfbox{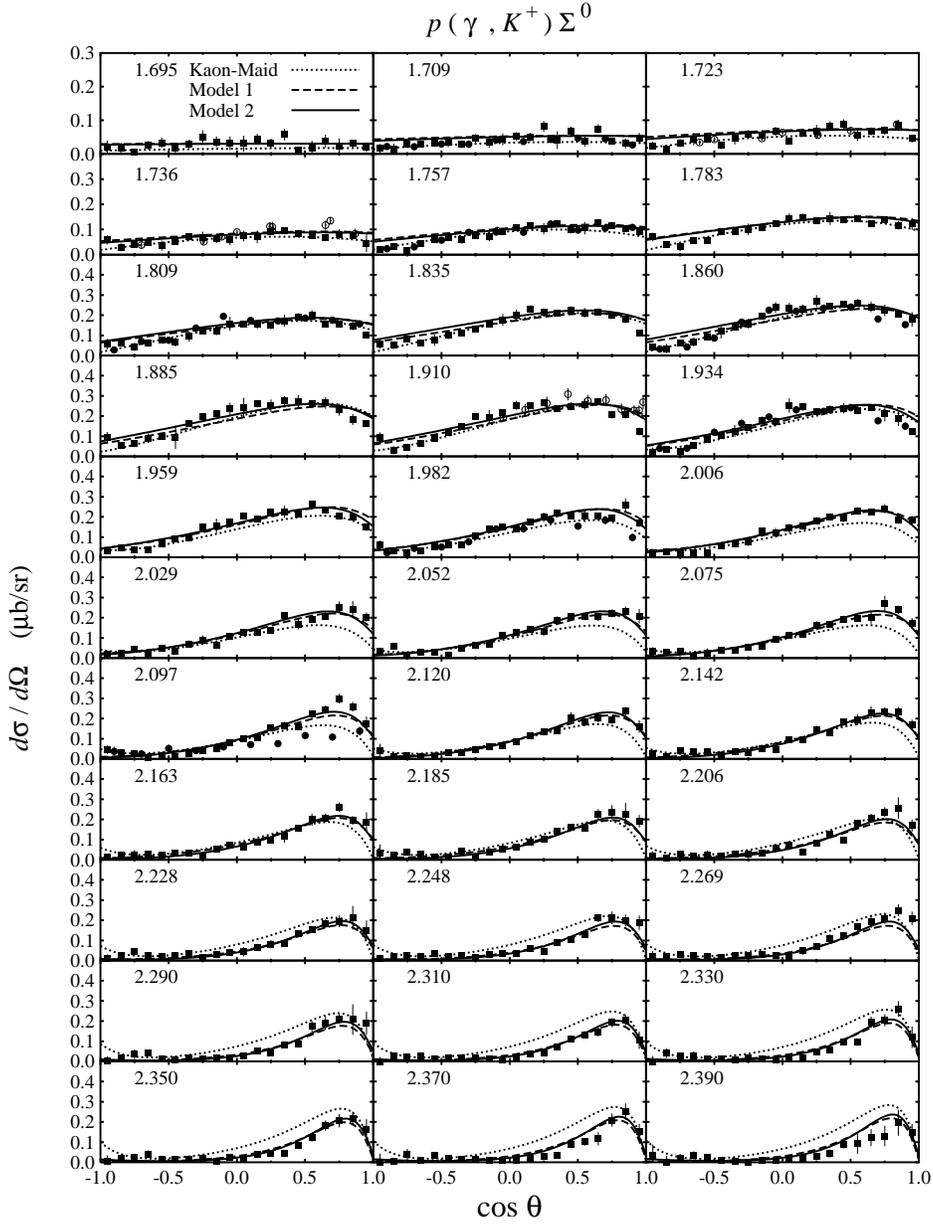}}   
\caption{Differential cross section for the $\gamma p\to K^+\Sigma^0$ channel.
  Experimental data are taken from Ref.\,\protect\cite{Glan03b}. Some older measurements
  are also shown for comparison.
 \label{kps0}}
\end{figure}

In the $K^+\Sigma^0$ channel the difference between the last two models does not
clearly show up, but this is not the case of the $K^0\Sigma^+$ channel. 
As shown by Fig.\,\ref{total}, it is obvious that Model 1 cannot reproduce the
$K^0\Sigma^+$ total cross section data at energies below 1900 MeV due to the lack of 
resonances with $M\approx 1800$ MeV. In Model 2 the fitted mass of the 
first $S_{11}$ and that of the $P_{31}$ are in this 
region. We also note that in the case of {\small KAON-MAID} the divergent behavior of 
the total cross section at high energies is attributed to the large value
of the hadronic form factor cut-off (0.82 GeV). At this region a slight increment in 
total cross section is also observed in Model 1, but not in Model 2.

Figure \ref{kps0} shows differential cross sections obtained from all 
models compared with experimental $K^+\Sigma^0$ data. In this case, similar
to the case of the total cross section, only {\small KAON-MAID} substantially
deviates from experimental data.

\begin{figure}[!t] 
\centerline{\epsfxsize=2.8in\epsfbox{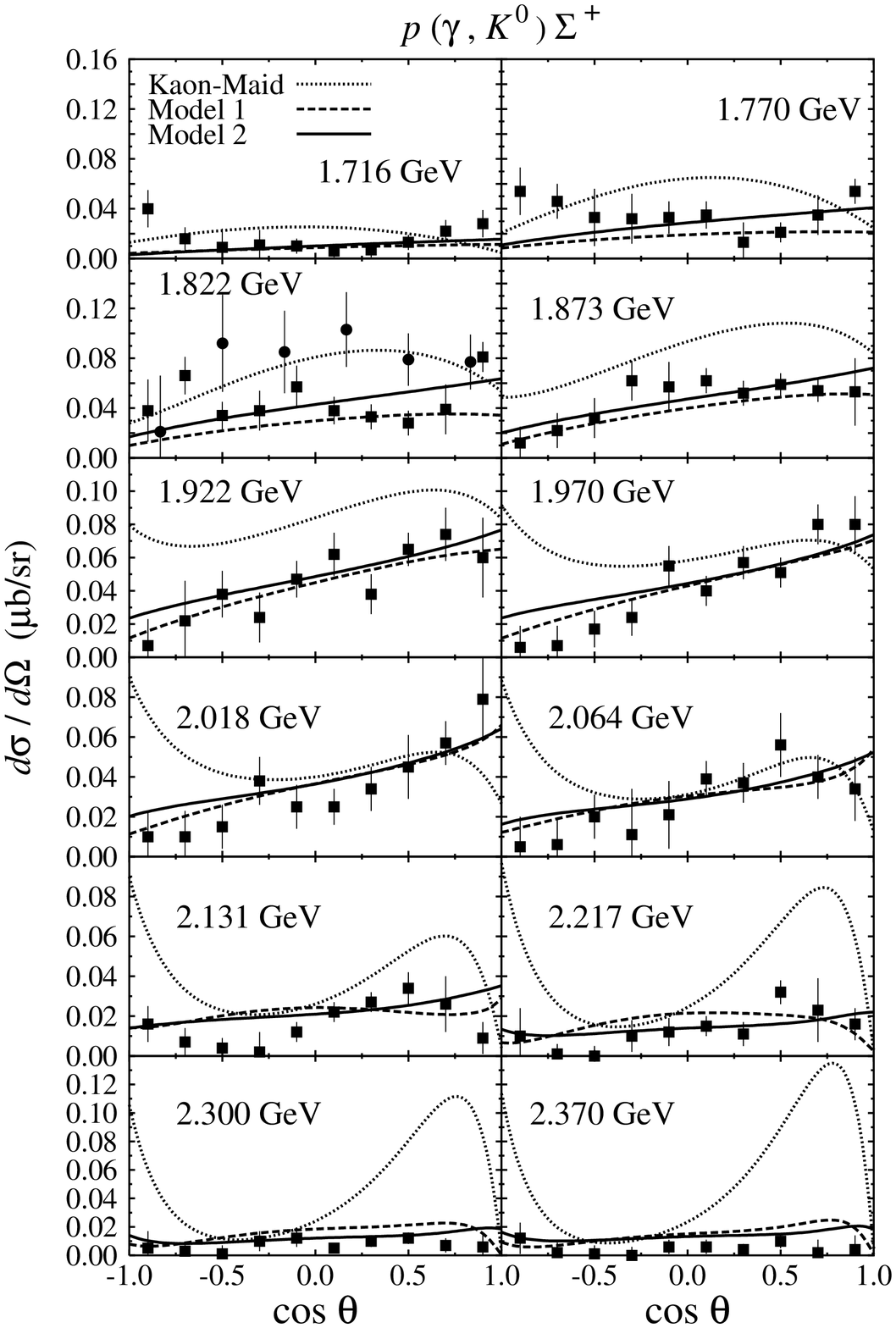}}   
\caption{Differential cross section for the $\gamma p\to K^0\Sigma^+$ channel.
  Experimental data are taken from Ref.\,\protect\cite{Lawall:2005np} (solid squares)
  and Ref.\,\protect\cite{Goers:1999sw} (solid circles).
 \label{k0sp}}
\end{figure}

Figure \ref{k0sp} compares the differential cross sections for the $K^0\Sigma^+$
channel. It is evident from this figure that KAON-MAID is
unable to reproduce the shape as well as the magnitude of differential cross
sections. In contrast to this, both Model 1 and Model 2 can fairly describe 
these new data up to some structures shown, e.g., at low energies and backward 
angles. The main difference from these models can be seen at forward directions,
where Model 1 tends to produce more forward peaking cross sections at high
energies.

\section{Conclusion}
We have shown that new {\small SAPHIR} data provide a stringent constrain to
isobar models such as {\small KAON MAID}. New resonances are required by
the models in order to explain the data. Whether or not they are ``missing 
resonances'' should be checked in the future by the coupled channels studies.

\end{document}